# 基於後量子密碼學的同態加密演算法

Abel C. H. Chen
Chunghwa Telecom Co., Ltd.
chchen.scholar@gmail.com; ORCID 0000-0003-3628-3033

## 摘要

隨著 Shor 演算法被提出，已證實量子電腦能在多項式時間(polynomial time)破解部分 NP 問題(如：質因數分解問題和離散對數問題)。近年來，雖然已有同態加密演算法被提出，但這些演算法可能是建構在基於質因數分解問題的密碼學基礎上，可能存在被量子計算破解的風險。有鑑於此，本研究提出基於後量子密碼學的同態加密演算法，在基於糾錯碼的密碼學上設計同態加密函數，提供可抵抗量子計算的同態加密演算法。在 3.2 節中提供數模型證明本研究方法的可行性，在 3.3 節提供實例說明本研究方法的作法。在實驗環境中，與現行主流的 RSA (Rivest-Shamir-Adleman)密碼學和橢圓曲線密碼學(Elliptic Curve Cryptography, ECC)進行比較，結果顯示加解密時間低於現行主流方法；同時，本研究提出的方法建構在非負矩陣分解問題(即 NP 問題)上，可以抵抗量子計算攻擊。

**關鍵字**：*後量子密碼學、同態加密、McEliece 密碼學*。

## 1. 前言

近年來，隨著量子電腦和量子演算法(如：Shor 因數分解演算法[1])的發展，逐漸造成 RSA (Rivest-Shamir-Adleman)密碼學和橢圓曲線密碼學(Elliptic Curve Cryptography, ECC)等非對稱金鑰密碼學安全上的威脅。其中，RSA 密碼學主要建構在質因數分解問題[2]，橢圓曲線密碼學主要建構在離散對數問題[3]，而質因數分解問題和離散對數問題分別為 NP 問題之一，從而支持 RSA 密碼學和橢圓曲線密碼學的安全性。然而，在 Shor 因數分解演算法被提出後，通過結合量子的特性，可以把質因數分解問題和離散對數問題等 NP 問題變成是多項式時間(polynomial time)可以解決的問題。有鑑於此，美國國家標準及技術研究所(National Institute of Standards and Technology, NIST)開始向全世界徵求後量子密碼學方法(post-quantum cryptography, PQC)，擬定新一代密碼學方法的標準[4]，以對抗量子計算的攻擊。

在後量子密碼學方法主要可以分為五大類，分別為：(1). 基於格的密碼學 (lattice-based cryptography)[5]、(2). 基於糾錯碼的密碼學(code-based cryptography)[6]、(3). 基於雜湊的密碼學(hash-based cryptography)[7]、(4). 基於多變量的密碼學(multivariate-based cryptography)[4]、(5). 基於超奇異橢圓曲線的密碼學 (supersingular elliptic curve-based cryptography)[4]。其中，由於基於格的密碼學在金鑰長度和加解密時間上優於其他密碼學方法，所以在美國國家標準及技術研究所徵選的第 3 輪技術報告中評選出了 CRYSTALS–Kyber 加解密演算法作為標準演算法。除此之外，為防範未來基於格的密碼學被破解，所以進行第 4 輪評選其他系列的密碼學方法作為備選標準。其中，基於糾錯碼的密碼學成為第 4 輪評選主要的候選加解密演算法，如：McEliece 密碼學方法。

雖然目前已有同態加密演算法被提出，但這些演算法可能是建構在 ElGamal 密碼學和基於質因數分解問題的密碼學基礎上，可能存在被量子計算破解的風險。有鑑於此，本研究以基於糾錯碼的密碼學為基礎，設計同態加密演算法，提出基於後量子密碼學的同態加密演算法，讓同態加密演算法具有抵抗量子計算的能力。本研究主要採用後量子密碼學方法之 McEliece 密碼學方法作為核心演算法

基礎，建構非負矩陣來進行同態加密和解密。由於非負矩陣分解問題是 NP 問題，並且目前量子計算尚無法破解非負矩陣分解問題，從而保障此方法可以抵抗量子計算。

本論文共分為五個章節。第 2 節中介紹 McEliece 密碼學的加解密方法。第 3 節提出本研究的基於後量子密碼學的同態加解密演算法，並且運用數學模型和提供實例說明本研究所提方法。第 4 節提供金鑰產製時間、加密時間、以及解密時間的效能比較。第 5 節總結本研究貢獻和討論未來研究方向。

## 2. McEliece 密碼學

McEliece 密碼學主要是建構在非負矩陣分解(non-negative matrix factorization, NMF)這個 NP 問題上的非對稱式金鑰密碼學[8]，以下從產製金鑰階段、加密階段、以及解密階段分別說明。

### 2.1 產製金鑰階段

產製金鑰階段步驟包含：

(1). 產生 3 個矩陣作為私鑰($S, G, P$)，其中，$S$ 作為擾碼器(scrambler)，為可逆矩陣(invertible matrix)。$G$ 為編碼生成器矩陣(encoding generator matrix)，有對應的奇偶檢定矩陣(parity-check matrix) $H$ 和解碼矩陣(decoding matrix) $R$，可用於糾錯(error correction)。$P$ 為置換矩陣(permutation matrix)。

(2). 將私鑰的 3 個矩陣相乘作為公鑰$\Psi$，如公式(1)所示。由於矩陣$\Psi$為非負矩陣，而非負矩陣分解是 NP 問題，從而保證 McEliece 密碼學的安全性。

$$\Psi = SGP \tag{1}$$

### 2.2 資料加密階段

資料加密階段步驟包含：

(1). 假設明文矩陣 $x$，並且產生隨機數矩陣 $e$。

(2). 通過公式(2)，以公鑰$\Psi$對明文 $x$ 加密，並有入隨機數 $e$ 擾動，產生密文 $y$。

$$y = x\Psi + e \tag{2}$$

### 2.3 資料解密階段

資料解密階段步驟包含：

(1). 通過公式(3)，以私鑰 $P$ 的反矩陣(即$P^{-1}$)對密文 $y$ 解密，取得$xSG + eP^{-1}$。

$$\begin{aligned}yP^{-1} &= (x\Psi + e)P^{-1} \\ &= xSGPP^{-1} + eP^{-1} \\ &= xSG + eP^{-1}\end{aligned} \tag{3}$$

(2). 通過公式(4)對矩陣$xSG + eP^{-1}$進行糾錯。根據矩陣$(xSG + eP^{-1})H$可以取得錯誤的位元位置，並進行更正。更正後可去除隨機數 $e$ 擾動的影響，也就是把矩陣 $xSG + eP^{-1}$更正為矩陣$xSG$。

$$yP^{-1}H = (xSG + eP^{-1})H \tag{4}$$

(3). 通過公式(5)，以私鑰 $G$ 對應的解碼矩陣 $R$ 對密文 $y$ 解密，取得$xS$。

$$xSGR = xS \tag{5}$$

(4). 通過公式(6)，以私鑰 $S$ 的反矩陣(即$S^{-1}$)對密文 $y$ 解密，取得明文 $x$。

$$xSS^{-1} = x \tag{6}$$

## 3. 研究方法

本研究主要提出基於後量子密碼學的同態加密演算法，可以在後量子密碼學基礎上提供同態加密應用。在 3.1 節中定義同態加密演算法的目標，在 3.2 節中描述本研究提出基於後量子密碼學的同態加密演算法及其數學模型證明，並在 3.3 節中提供實例證明本研究提出的基於後量子密碼學的同態加密演算法。3.4 節小結方法特色和討論其限制。

## 3.1 同態加密演算法的目標

本研究主要從事半同態加密 (partially homomorphic encryption)的研究，並主要進行加法同態加密(additive homomorphic encryption)，所以本節定義加法同態加密演算法的目標。其中，本節假設明文為$x_i$，通過加法同態加密函數$Enc(x_i)$計算後可以得到密文$c_{x_i}$，如公式(7)所示；並且有加法同態解密函數$Dec(c_{x_i})$可對密文$c_{x_i}$進行解密後得到$x_i$，如公式(8)所示。

$$c_{x_i} = Enc(x_i) \tag{7}$$

$$x_i = Dec(c_{x_i}) \tag{8}$$

由於本研究主要考量二進制矩陣(binary matrix)運算，所以加法運算採用邏輯互斥或(XOR)計算，並以符號$\oplus$表示。公式(9)和(10)是加法同態加密目標，密文加總後的結果解密後會等於明文加總後的結果。因此，加法同態加密函數$Enc(x_i)$和加法同態解密函數$Dec(c_{x_i})$需同時符合公式(9)和(10)的特性。

$$\Phi = \sum_{i=1}^{n} c_{x_i} = c_{x_1} \oplus c_{x_2} \oplus ... \oplus c_{x_n} \tag{9}$$

$$\Phi = Enc(\Theta), \text{ where } \Theta = \sum_{i=1}^{n} x_i = x_1 \oplus x_2 \oplus ... \oplus x_n \tag{10}$$

## 3.2 基於後量子密碼學的同態加密演算法及原理

本節介紹本研究提出的後量子密碼學的同態加密演算法，主要結合 McEliece 密碼學的特性建構邏輯互斥或(XOR)的加法同態加密演算法。其中，假設私鑰為(**S**, **G**, **P**)、公鑰為$\Psi$ (通過公式(1)計算得到)。本研究修改 McEliece 密碼學的加密函數，設計本研究提出的加法同態加密函數$Enc(x_i)$如公式(11)所示，避免隨機數 **e** 擾動，並且仍保持非負矩陣特性，以符合安全性(非負矩陣分解是 NP 問題)[9]。

$$c_{x_i} = Enc(x_i) = x_i \Psi \tag{11}$$

通過公式(12)可以證明本研究所設計的加法同態加密函數$Enc(x_i)$可以達到加法同態加密演算法的目標(即符合公式(9)和(10)的特性)。在解密時，可以採用傳統的 McEliece 密碼學的解密過程即可解密。

$$\begin{aligned}
\Phi &= \sum_{i=1}^{n}(c_{x_i}) \\
&= c_{x_1} \oplus c_{x_2} \oplus ... \oplus c_{x_n} \\
&= x_1\Psi \oplus x_2\Psi \oplus ... \oplus x_n\Psi \\
&= (x_1 \oplus x_2 \oplus ... \oplus x_n)\Psi \\
&= \Theta\Psi = Enc(\Theta)
\end{aligned} \tag{12}$$

## 3.3 實例說明

本節以實例說明後量子密碼學的同態加密演算法，在 3.3.1 節中說明本實例產製的金鑰矩陣，在 3.3.2 節中以 2 筆資料明文以加法同態加密函數$Enc(x_i)$進行加密，3.3.3 節以加法同態解密函數$Dec(c_{x_i})$進行解密。

### 3.3.1 產製金鑰階段

本實例採用的私鑰為(**S**, **G**, **P**)，分別如公式(13)、(14)、(15)所示，並根據公式(1)產生公鑰$\Psi$，如公式(16)所示。其中，在糾錯碼方法採用漢明碼(hamming code)方法，產生編碼生成器矩陣 **G** 對應的奇偶檢定矩陣 **H** 和解碼矩陣 **R**，分別如公式(17)、(18)所示。後續可通過這些矩陣進行加密和解密計算。

$$\mathbf{S} = \begin{bmatrix} 1 & 1 & 0 & 1 \\ 1 & 0 & 0 & 1 \\ 0 & 1 & 1 & 1 \\ 1 & 1 & 0 & 0 \end{bmatrix} \tag{13}$$

$$\mathbf{G} = \begin{bmatrix} 1 & 1 & 1 & 0 & 0 & 0 & 0 \\ 1 & 0 & 0 & 1 & 1 & 0 & 0 \\ 0 & 1 & 0 & 1 & 0 & 1 & 0 \\ 1 & 1 & 0 & 1 & 0 & 0 & 1 \end{bmatrix} \quad (14)$$

$$\mathbf{P} = \begin{bmatrix} 0 & 1 & 0 & 0 & 0 & 0 & 0 \\ 0 & 0 & 0 & 1 & 0 & 0 & 0 \\ 0 & 0 & 0 & 0 & 0 & 0 & 1 \\ 1 & 0 & 0 & 0 & 0 & 0 & 0 \\ 0 & 0 & 1 & 0 & 0 & 0 & 0 \\ 0 & 0 & 0 & 0 & 0 & 1 & 0 \\ 0 & 0 & 0 & 0 & 1 & 0 & 0 \end{bmatrix} \quad (15)$$

$$\mathbf{\Psi} = \mathbf{SGP} = \begin{bmatrix} 0 & 1 & 1 & 0 & 1 & 0 & 1 \\ 1 & 0 & 0 & 0 & 1 & 0 & 1 \\ 1 & 0 & 1 & 0 & 1 & 1 & 0 \\ 1 & 0 & 1 & 1 & 0 & 0 & 1 \end{bmatrix} \quad (16)$$

$$\mathbf{H} = \begin{bmatrix} 1 & 0 & 0 \\ 0 & 1 & 0 \\ 1 & 1 & 0 \\ 0 & 0 & 1 \\ 1 & 0 & 1 \\ 0 & 1 & 1 \\ 1 & 1 & 1 \end{bmatrix} \quad (17)$$

$$\mathbf{R} = \begin{bmatrix} 0 & 0 & 0 & 0 \\ 0 & 0 & 0 & 0 \\ 1 & 0 & 0 & 0 \\ 0 & 0 & 0 & 0 \\ 0 & 1 & 0 & 0 \\ 0 & 0 & 1 & 0 \\ 0 & 0 & 0 & 1 \end{bmatrix} \quad (18)$$

### 3.3.2 資料加密階段

假設有 2 筆資料明文(即 $n = 2$)分別為 $x_1 = [0\ 1\ 0\ 0]$、$x_2 = [1\ 0\ 0\ 0]$，分別通過公式(11)進行加密，如公式(19)、(20)所示。

$$c_{x_1} = Enc(x_1) \\ = x_1 \mathbf{\Psi} = [1\ 0\ 0\ 0\ 1\ 0\ 1] \quad (19)$$

$$c_{x_2} = Enc(x_2) \\ = x_2 \mathbf{\Psi} = [0\ 1\ 1\ 0\ 1\ 0\ 1] \quad (20)$$

在加法同態加密的應用上，可對密文進行加總，並且密文加總後的結果解密後可以得到明文加總後的結果。本實例密文加總後的結果如公式(21)所示。

$$\Phi = \sum_{i=1}^{2}(c_{x_i}) \\ = c_{x_1} \oplus c_{x_2} \\ = [1\ 0\ 0\ 0\ 1\ 0\ 1] \oplus \\ [0\ 1\ 1\ 0\ 1\ 0\ 1] \\ = [1\ 1\ 1\ 0\ 0\ 0\ 0] \quad (21)$$

### 3.3.3 資料解密階段

本節對密文加總後的結果 $[1\ 1\ 1\ 0\ 0\ 0\ 0]$ 進行解密，如公式(22)所示。並且，通過公式(23)可以驗證密文加總後的結果解密後可以等於明文加總後的結果。

$$Dec(\Phi) = \Phi \mathbf{P}^{-1} \mathbf{R} \mathbf{S}^{-1} = [1\ 1\ 0\ 0] \quad (22)$$

$$\Theta = \sum_{i=1}^{2} x_i = x_1 \oplus x_2 \\ = [0\ 1\ 0\ 0] \oplus [1\ 0\ 0\ 0] \\ = [1\ 1\ 0\ 0] = Dec(\Phi) \quad (23)$$

## 3.4 小結與討論

通過3.2節以數學模型證明本研究提出的基於後量子密碼學的同態加密演算法可行性，並通過3.3節以實例說明本研究提出的基於後量子密碼學的同態加密演算法。未來可以將此方法應用在密文情況下進行邏輯互斥或(XOR)計算的應用。

由於本研究提出的基於後量子密碼學的同態加密演算法限制在邏輯互斥或(XOR)的計算，未來可嘗試往一般的加法和乘法上改進。

## 4. 實證比較與討論

為實際驗證本研究提出的基於後量子密碼學的同態加密演算法之效率，本節與現行主流的RSA (Rivest-Shamir-Adleman)密碼學和橢圓曲線密碼學

(Elliptic Curve Cryptography, ECC)進行比較。在實驗環境中，本研究使用一台 Windows 10 企業版的電腦執行演算法，分別驗證資料加密時間(如 4.1 節)和資料解密時間(如 4.2 節)。其中，實驗環境中使用到的軟硬體包含有 CPU Intel(R) Core(TM) i7-10510U、記憶體 8 GB、OpenJDK 18.0.2.1、以及函式庫 Bouncy Castle Release 1.72。

**4.1 資料加密時間**

本節分別採用不同的資料筆數來比較各個方法的資料加密時間，包含 100 筆、1,000 筆、10,000 筆、以及 100,000 筆。本研究方法採用公式(11)對資料進行加密；在資料筆數 100 筆時(即 $n=100$)，本研究方法總運算時間為 0.131 毫秒。在同樣資料量為 100 筆時，RSA 資料加密的總運算時間為 1.778 毫秒，ECC 資料加密的總運算時間為 686.517 毫秒。並且，隨著資料量的增加，運算時間將呈現倍數成長。其中，由於橢圓曲線密碼學在加密的計算上採用橢圓曲線整合加密機制(Elliptic Curve Integrated Encryption Scheme, ECIES)，在過程中需要先計算橢圓曲線 Diffie–Hellman 金鑰交換(Elliptic Curve Diffie–Hellman key exchange, ECDH)再取得共享金鑰的橢圓曲線點，並根據橢圓曲線點產生進階加密標準(Advanced Encryption Standard, AES)金鑰和簽章金鑰，再用進階加密標準演算法進行加密，所以需要大量的運算時間。

表 1 資料加密時間(單位：毫秒)

| 資料筆數 | RSA | ECC | 本方法 |
| --- | --- | --- | --- |
| 100 | 1.778 | 686.517 | **0.131** |
| 1,000 | 9.724 | 6681.865 | **0.848** |
| 10,000 | 32.581 | 59449.652 | **6.551** |
| 100,000 | 125.750 | 421820.914 | **28.221** |

**4.2 資料解密時間**

本節對 4.1 節的資料進行同態解密，驗證解密時間。其中，由於本研究對比方法 RSA 和 ECC 不具備同態加解密函數，所以在解密驗證上採用對逐筆資料解密後加總的時間。在本研究方法的驗證上，同態解密採用公式(12)計算，並在密文加總後做一次解密計算。由實驗結果顯示，在資料筆數 100 筆時(即 $n=100$)，本研究方法總運算時間為 0.058 毫秒。在同樣資料量為 100 筆時，RSA 資料加密的總運算時間為 1.928 毫秒，ECC 資料加密的總運算時間為 619.556 毫秒。並且，隨著資料量的增加，運算時間將呈現倍數成長。其中，由於橢圓曲線密碼學在解密的計算上採用橢圓曲線整合加密機制，所以需要較多的運算量，所需要的運算時間也較多。

表 2 資料解密時間(單位：毫秒)

| 資料筆數 | RSA | ECC | 本方法 |
| --- | --- | --- | --- |
| 100 | 1.928 | 619.556 | **0.058** |
| 1,000 | 11.183 | 6694.206 | **0.467** |
| 10,000 | 40.203 | 59983.918 | **4.230** |
| 100,000 | 148.471 | 427265.235 | **16.620** |

**5. 結論與未來研究**

本研究提出基於後量子密碼學的同態加密演算法，在基於糾錯碼的後量子密碼學方法(即 McEliece 密碼學方法)基礎上設計同態加密函數，讓同態加密演算法具有抵抗量子計算攻擊的能力。本研究分別從數學模型和實例來論證提出的基於後量子密碼學的同態加密演算法可行性。在實驗結果中顯示本研究方法的加密效率和解密效率高於現行主流的 RSA 和 ECC 演算法，並且可以避免量子攻擊。未來可嘗試不同安全強度下的驗證。

# Homomorphic Encryption Based on Post-Quantum Cryptography

Abel C. H. Chen

Chunghwa Telecom Co., Ltd.

chchen.scholar@gmail.com

**Abstrac**t

With the development of Shor's algorithm, some nondeterministic polynomial (NP) time problems (e.g. prime factorization problems and discrete logarithm problems) may be solved in polynomial time. In recent years, although some homomorphic encryption algorithms have been proposed based on prime factorization problems, the algorithms may be cracked by quantum computing attacks. Therefore, this study proposes a post-quantum cryptography (PQC)-based homomorphic encryption method which includes the homomorphic encryption function based on a code-based cryptography method for avoiding quantum computing attacks. Subsection 3.2 proposes mathematical models to prove the feasibility of the proposed method, and Subsection 3.3 gives calculation examples to present the detailed steps of the proposed method. In experimental environments, the mainstream cryptography methods (i.e. RSA cryptography and elliptic curve cryptography (ECC)) have been compared, and the results show that the encryption time and decryption time of the proposed method are shorter than other cryptography methods. Furthermore, the proposed method is designed based on a non-negative matrix factorization problem (i.e. a NP problem) for resisting quantum computing attacks.

***Keywords***: *Post-quantum cryptography, homomorphic encryption, McEliece cryptography*